\documentclass[12pt,preprint]{emulateapj}

\shorttitle{Rotation and Stellar Populations}
\shortauthors{Levesque et al.}

\begin{document}

\title{The Effects of Stellar Rotation. I. Impact on the Ionizing Spectra and Integrated Properties of Stellar Populations}
\author{Emily M. Levesque$^{1,5}$, Claus Leitherer$^{2}$, Sylvia Ekstrom$^{3}$, Georges Meynet$^{3}$, Daniel Schaerer$^{3,4}$}

\begin{abstract}
We present a sample of synthetic massive stellar populations created using the Starburst99 evolutionary synthesis code and new sets of stellar evolutionary tracks, including one set that adopts a detailed treatment of rotation. Using the outputs of the Starburst99 code, we compare the populations' integrated properties, including ionizing radiation fields, bolometric luminosities, and colors. With these comparisons we are able to probe the specific effects of rotation on the properties of a stellar population. We find that a population of rotating stars produces a much harder ionizing radiation field and a higher bolometric luminosity, changes that are primarily attributable to the effects of rotational mixing on the lifetimes, luminosities, effective temperatures, and mass loss rates of massive stars. We consider the implications of the profound effects that rotation can have on a stellar population, and discuss the importance of refining stellar evolutionary models for future work in the study of extragalactic, and particularly high-redshift, stellar populations.
\end{abstract}

\section{Introduction}
\footnotetext[1]{CASA, Department of Astrophysical and Planetary Sciences, University of Colorado 389-UCB, Boulder, CO 80309, USA; \texttt{Emily.Levesque@colorado.edu}}
\footnotetext[2]{Space Telescope Science Institute, 3700 San Martin Drive, Baltimore, MD 21218, USA}
\footnotetext[3]{Geneva Observatory, University of Geneva, Maillettes 51, CH-1290 Sauverny, Switzerland}
\footnotetext[4]{CNRS, IRAP, 14 Avenue E. Belin, 31400 Toulouse, France}
\footnotetext[5]{Einstein Fellow}

Stellar populations are the key component driving the radiative properties of star-forming galaxies. In the absence of an active galactic nucleus, the young massive star population is the primary source of ionizing radiation (e.g. Baldwin et al.\ 1981, Gonz\'{a}lez Delgado \& Leitherer 1999). The integrated light of underlying older stellar populations will typically dominate the galaxy's continuum emission, and can be used to probe the galaxy's stellar mass and star formation history (e.g. Cid Fernandes et al.\ 2001; Gonz\'{a}lez Delgado et al.\ 1999, 2005; Kauffmann et al.\ 2003a,b). Finally, at large lookback time stellar populations are vital in studies of early star formation and its impact on galaxy evolution, the initial mass function, and reionization (e.g. Shull et al.\ 2011).

Evolutionary synthesis codes are used to simulate the integrated properties of stellar populations. The models produced by these codes are widely used to interpret observations of galaxies. Such models are vital in studies of galaxies' star formation histories and their evolution with time. In conjunction with photoionization codes, the SEDs produced by evolutionary spectral synthesis can also be used to generate model galaxy spectra that can be applied to observations of star-forming galaxies, probing their interstellar medium properties and star formation histories (e.g. Kewley et al.\ 2001, Moy et al.\ 2001, Fernandes et al.\ 2003, Dopita et al.\ 2006). These same model spectra are commonly used for calibrating interstellar medium diagnostics in the UV, optical, and infrared, making them a critical component in determinations of galaxy metallicities, ionization parameters, star formation rates, and star formation histories (e.g. Kewley \& Dopita 2002, Kobulnicky \& Kewley 2004, Tremonti et al.\ 2004, Rix et al.\ 2004, Snijders et al.\ 2007). Even cosmological studies, which depend on a solid theoretical framework that can be used to determine galaxies' luminosities, M/L ratios, and chemical evolution, are ultimately dependent on these model spectra.

Evolutionary synthesis codes are in turn dependent upon stellar evolutionary models. These models face a number of challenges. Prior generations of these models (Schaller et al.\ 1992, Meynet et al.\ 1994, Bressan et al.\ 1993), which are still currently used in many evolutionary synthesis codes, included only limited treatments of phenomena such as stellar mixing and mass loss, binarity, and rotation. However, in recent years improvements have been made on both the observational and theoretical sides. Stellar evolutionary models have begun to consider different mass loss rates, the effects of binarity, and treatments of rotation in detail (e.g. Meynet et al.\ 1994, Maeder \& Meynet 2000, de Mink et al.\ 2009, V\'{a}zquez et al.\ 2007). New observations of stellar populations both within the Milky Way and across the Local Group have also offered a rich comparison sample that can be used to calibrate evolutionary models across a range of initial masses and metallicities (e.g. Villamariz et al.\ 2002, Levesque et al.\ 2005, 2006, Massey et al.\ 2006, 2007, Searle et al.\ 2008, Przybilla et al.\ 2010).

Recently, Ekstr\"{o}m et al.\ (2012) published two new grids of stellar evolutionary tracks for single stars at solar metallicity produced by the Geneva group. These tracks include a number of updates from the previous solar-metallicity grids published by Schaller et al.\ (1992) and Meynet et al.\ (1994), with changes made to physical ingredients such as abundances, nuclear reaction rates, and convection modeling. In addition, one of the grids adopts a detailed treatment of axial rotation effects on stellar evolution spanning the full mass range of the tracks (hereafter the ROT tracks). The stars in this model grid were given initial (zero-age main sequence) rotation rates of $v_{ini} = 0.4v_{crit}$, corresponding to the peak of the velocity distribution observed in young B stars (Huang et al.\ 2010). Adopting this constant ratio across the full range of evolutionary tracks rather than a single constant value for $v_{ini}$ ensured that rotation effects, which are dependent on this ratio, would be consistent across tracks of different masses. In addition to the ROT tracks, this same work also produces an otherwise-identical grid of evolutionary tracks that include no treatment of rotation (hereafter the NOROT tracks), allowing direct comparisons between the two grids that can isolate any changes specifically attributable to rotation. For a complete discussion of the rotation treatment and other physical parameters adopted by the Geneva models, see Ekstr\"{o}m et al.\ (2012) and references therein. 

The Ekstr\"{o}m et al.\ (2012) ROT tracks show a number of improvements compared to non-rotating evolutionary models. These tracks better reproduce the observed width of the lower-mass main sequence from Wolff \& Simon (1997) and the observed maximum luminosity for red supergiants (Levesque et al.\ 2005, 2006; Massey et al.\ 2009). In addition, the ROT tracks show a substantial improvement over the NOROT tracks at reproducing the observed populations of Wolf-Rayet (WR) stars (Georgy et al.\ 2012), and at correctly predicting the properties of observed yellow supergiant supernova progenitors (Georgy 2012). Similar results were also found when comparing ROT tracks to evolved supergiants in the Large Magellanic Cloud and M33 (Neugent et al.\ 2012, Drout et al.\ 2012). However, despite these advances the ROT tracks do still present several shortcomings. Georgy et al.\ (2012) note that these new tracks still struggle to properly reproduce the duration of different Wolf-Rayet phases. It also remains to be seen how well the ROT tracks reproduce observations of local stellar clusters. 

In this work we specifically examine the effects of rotation on the integrated properties of a stellar population. We use the Starburst99 evolutionary synthesis code to produce several model stellar populations using different evolutionary tracks, carefully considering how the newer Ekstr\"{o}m et al.\ (2012) tracks might influence the default input parameters adopted in the code (Section 2). We then compare the resulting stellar samples and several integrated properties of these populations, noting the specific changes caused by the inclusion of rotation (Section 3). Finally, we consider the implications of rotation's profound effects on a stellar population, noting existing shortcomings in the stellar evolutionary models and highlighting the impact that these new advances in stellar evolutionary modeling will have on larger studies that utilize models of stellar populations (Section 4).

\section{Starburst99 Parameters}
Starburst99 (Leitherer et al.\ 1999, 2010; Leitherer \& Chen 2009; V\'{a}zquez \& Leitherer 2005), is an evolutionary synthesis code that can be used to generate a synthetic stellar population and track integrated properties of the population such as ionization spectra, M/L, and colors. The populations are produced by combining model stellar atmospheres and spectra with grids of evolutionary tracks for massive stars. For this work, we have used Starburst99 to generate synthetic stellar populations using both the NOROT and ROT stellar evolutionary tracks from Ekstr\"{o}m et al.\ (2012). We have compared the resulting outputs both to each other and to the older Starburst99 outputs from Levesque et al.\ (2010), which adopt the non-rotating evolutionary tracks of Meynet et al.\ (1994; hereafter the Meynet94 tracks) that include enhanced mass-loss rates to roughly approximate rotation effects.

All of these models are computed with a solar chemical composition that corresponds to the metallicity adopted by the evolutionary tracks ($z = 0.02$ for the Meynet94 tracks, $z = 0.014$ for the NOROT and ROT tracks). All three model populations adopt a Salpeter initial mass function (IMF; Salpeter 1955) with $\alpha = 2.35$ and a 0.1$M_{\odot}$-100$M_{\odot}$ mass range. We simulate a zero-age instantaneous burst star formation history with a fixed mass of 10$^6 M_{\odot}$. Starting with this population at an initial 0 Myr age, we simulate the evolution of this single stellar population up to 10 Myr in 0.5 Myr increments. This means of modeling an instantaneous burst of star formation allows us to trace the effects of a single coeval stellar population and isolate the contributions of different stellar samples as the population ages. By modeling a fully sampled IMF we are able to examine the differential effects of rotation; however, it should be noted that stochastic effects are also expected to present in such samples. A detailed discussion of these effects is beyond the scope of this paper (see, for example, Cervi\~{n}o et al.\ 2000, Cervi\~{n}o \& Luridiana 2006, Fumagalli, et al.\ 2011, da Silva et al.\ 2012).

Starburst99 adopts model stellar atmospheres and spectra in conjunction with the evolutionary tracks. The model atmospheres available in the Starburst99 code include the static Kurucz models compiled in Lejeune et al.\ (1997), useful for approximating plane-parallel atmospheres, the Hillier \& Miller (1998) extended atmosphere models generated with CMFGEN, and the Pauldrach et al.\ (2001) WM-Basic extended atmospheres best used for O stars. The choice of different atmosphere models impacts the way in which surface temperatures are calculated for the model stellar populations. For plane-parallel models Starburst99 will adopt the surface temperatures given directly by the evolutionary tracks. However, for extended models Starburst99 will use radiative transfer treatments of the atmosphere models and core temperatures given by the evolutionary tracks to calculate a surface temperature. However, Smith et al.\ (2002) introduced a ``weighting factor" into the Starburst99 code, resulting in a mix of 60\% calculated surface temperatures (determined from the tracks' core temperatures) and 40\% direct surface temperatures (taken directly from the tracks' reported surface temperatures). This adjustment was meant to bring Starburst99 models calculated with the previous generation of Geneva evolutionary tracks (e.g. Schaller et al.\ 1992, Meynet et al.\ 1994) into agreement with empirical observations of FUV spectra. Here we consider the effect that modifying this percentile has on the Starburst99 outputs.

In Figure 1 we compare the number of H I, He I, and He II ionizing photons produced by the Starburst99 code as a function of time when adopting different atmosphere treatments and the ROT tracks. These include the Hillier \& Miller (1998) and Pauldrach et al.\ (2001) atmosphere models with weighting factors of 0.1 (adopting 10\% core temperatures), 1.0 (adopting 100\% core temperatures), and the current default of 0.6 (60\% core temperatures), as well as the traditional Lejeune et al.\ (1997) plane-parallel models. We note immediately that the apparent difference between these different temperature treatments is relatively minor when considering the number of H I ionizing photons, but becomes progressively more drastic as we move to shorter and more temperature-sensitive wavelengths, with a higher percentile of core temperatures corresponding to a harder spectrum for the extended atmosphere models. For the remainder of this work we adopt the default Starburst99 treatment for all of our synthetic stellar populations, using the extended model atmospheres with a weighting factor of 0.6. However, it is clear that changes to this weighting factor could impact the ionizing spectra produced by a stellar population, with stronger and more age-dependent effects becoming apparent at shorter wavelengths.

Finally, Starburst99 typically takes the definition of a WR star to be any star with log($T_{\rm eff}$) $>$ 4.4 and a surface hydrogen abundances of $<$40\% (e.g. Meynet \& Maeder 2003). However, we also consider the impact that altering these definitions would  have on the Starburst99 outputs, adopting a wider range of minimum temperatures (ranging from log($T_{\rm eff}$) $>$ 4.0 to log($T_{\rm eff}$) $>$ 4.8) and a lower surface hydrogen abundance of $<$10\%. This definition will affect when the atmosphere model that Starburst99 adopts for a given star shifts from the Pauldrach et al.\ (2001) models (more suitable for blue supergiants) to the Hillier \& Miller (1998) models (more suitable for Wolf-Rayet stars). From our comparison in Figure 2 adopting the ROT tracks, we see that changes to the WR star definition largely yield only small differences ($\le 0.3$ dex) in the number of H I, He I, and He II ionizing photons produced by Starburst99; the one exception is a significantly increased production of He II ionizing photons when the minimum temperature is raised to log($T_{\rm eff}$) $>$ 4.8, an apparent consequence of applying the Pauldrach et al.\ (2001) models at higher temperatures. As the definitions with lower minimum temperatures show only small variations and are in good agreement with observational data on Wolf-Rayet stars (e.g. Massey 2003, Crowther 2007), we adopt the default Starburst99 definition of a Wolf-Rayet star for the remainder of this work.

\section{Stellar Rotation Effects}
\subsection{Massive Star Populations}
In Figure 3 we compare the contribution of each individual stellar mass range to the H I ionizing continuum as a function of time, modeled by the NOROT (top) and ROT (bottom) evolutionary tracks. From this comparison we can see that the ROT stellar populations contribute to the H I ionizing continuum out to later ages, and that this difference becomes more pronounced in the lower-mass bins (e.g. the 80-100M$_{\odot}$ mass bin shows a cutoff in H I ionizing photon production that is 0.7 Myr later in the ROT models, while the 20-40M$_{\odot}$ mass bin shows a cutoff that is 2.0 Myr later in the ROT models). Stellar populations modeled with rotation also produce a greater number of H I ionizing photons at later ages.

This fundamental difference between the two models is due to the impact of rotational mixing on stellar lifetimes - stars will spend more time on the main sequence (for more discussion see Ekstr\"{o}m et al.\ 2012 and references therein), and will have both higher effective temperatures {\it and} higher luminosities for a given temperature than their non-rotating counterparts due to the effects of He enhancement (Leitherer 2008, Leitherer \& Ekstr\"{o}m 2012). Rotational mixing also leads to stronger mass loss, favoring a blueward evolution for stars in their red supergiant phase (Salasnich et al.\ 1999) and lowering the minimum initial mass for a WR star from 32$M_{\odot}$ in the NOROT tracks to 20$M_{\odot}$ in the ROT tracks. This in turn greatly increases the range of ages for WR stars, resulting in a single stellar population having a larger and longer-lived WR sample (Georgy et al.\ 2012). The net result of these phenomena is that a single coeval population of rotating stars will produce more hot massive stars at both early and later ages, therefore generating a H I ionizing continuum with a greater number of photons {\it and} a considerably longer lifetime. 

\subsection{Ionizing Spectra}
In the previous section we adopted the number of H I, He I, and He II ionizing photons as proxies for the hardness of the ionizing radiation spectrum. Here we directly examine the far-ultraviolet (FUV) spectrum produced by the total stellar population. Figure 4 plots the synthetic FUV spectrum produced by Starburst99 for our ROT, NOROT, and Meynet94 synthetic stellar populations, comparing the spectra as the population ages from 1 to 10 Myr. 

It is immediately apparent from both of these figures that the FUV spectrum produced by the ROT population is considerably harder than those produced by either the NOROT or Meynet94 stellar populations. This difference is particularly pronounced at wavelengths shorter than $\sim$228\AA, the ionization threshold of He II, and at later ages, from 7-10 Myr, corresponding to an increase in flux of at least $\sim$10 dex for the ROT population at $\lesssim$228\AA\ and as much as $\sim$7 dex at $\gtrsim$228\AA. Levesque et al.\ (2010) noted the appearance of a ``bump" in the high-energy region of the Meynet94 spectrum; the feature appears blueward of $\sim$195\AA\ (spanning the CIV, OIV, and NV ionizing thresholds) in the 3-5Myr Meynet94 spectra, corresponding with the age range of Wolf-Rayet stars in the Meynet94 tracks. This same feature is also apparent in the NOROT and ROT spectra; however, while it is again present from 3-5 Myr in the NOROT spectra, it is discernible up through 9 Myr in the ROT models. This corresponds to the increased range of Wolf-Rayet initial masses (and therefore ages) in a rotating stellar population, dramatically increasing the high-energy FUV flux produced by the ROT models. In addition, there is a notable and unique increase in both the flux and high-energy limit of the ROT FUV spectrum at 7 Myr; this is explained by the large number of hot Wolf-Rayet stars at lower masses, and therefore later ages, that is unique to the ROT population. 

Similarly, we also consider the ratio of the He II to H I continuua as a function of age (Figure 5), another tracer of hardness in the ionizing spectrum. From Figure 4 we can see that, for all three populations, the H I continuum flux stays constant from 1 to 10Myr while the He II continuum flux drops sharply with age. As a result, this diagnostic acts as a good tracer of the time-dependent behavior of the He II  continuum. We can see that there is a sharp decrease in the He II flux for all three models; however, while these drops are comparable in scale, the decrease happens significantly earlier for the NOROT and Meynet94 populations (at 3 Myr) than in the ROT population (at 5 Myr), corresponding to the effects of rotation on the lifetimes and populations of hot massive stars that can sustain a higher He II continuum flux.

Combined, our examination of these ionizing spectra highlights the enormous impact that stellar rotation can have on a massive star population and its role as the primary ionizing source in a star-forming galaxy. A rotating stellar population produces far more high-energy photons and maintains this harder FUV spectrum for $\sim$2 Myr longer than a non-rotating population. This hardening of the spectrum goes in the right direction; Levesque et al.\ (2010) found that the FUV spectrum produced by the Meynet94 models was not hard enough to reproduce observations of emission line fluxes in star-forming galaxies, and suggested that rotating models might be one way of improving this discrepancy. However, the current differences in the models are quite extreme. The hardness of the ionizing radiation field (measured by the He II/H I ratio and a key component in production of emission lines and other properties driven by a stellar population) changes by up to $\sim$3 dex when comparing the NOROT and ROT models, showing that the ROT population produces a much harder ionizing radiation field and a substantially increased flux of ionizing photons. While it remains to be seen whether this predicted increase agrees with the properties of observed stellar populations, these results nevertheless highlight the importance of including stellar rotation treatments in models of stellar populations.

\subsection{Bolometric Luminosity}
In addition to its impact on the ionizing spectrum produced by a stellar population, rotation also directly impacts the population's bolometric luminosity ($M_{\rm bol}$). Comparing the evolution of $M_{\rm bol}$ with age for our three models (Figure 6), we can see that the population modeled with rotation is $\gtrsim$0.4 mag more luminous than the newly-modeled non-rotating population from $\sim$4 Myr onward (see also Leitherer \& Ekstr\"{o}m 2012 and the earlier results of V\'{a}zquez et al.\ 2007). This increase in luminosity extends all the way out to 10 Myr, past the predicted age of the Wolf-Rayet phase; this demonstrates that a substantial contribution to the increased luminosity of the sample is produced by main-sequence late-type O stars. Combined with the results of Ekstr\"{o}m et al.\ (2012), it is clear that O stars in the ROT models are nearly 50\% more luminous at a given temperature. This corresponds to larger predicted radii for these stars, and a decrease in the modeled mass-to-light ratio of star-forming galaxies, impacting derivations of the IMF for a rotating population.

\subsection{Colors}
In Figure 7 we compare the evolution of the $U-B$, $B-V$, $V-R$, and $V-K$ indices for the ROT, NOROT, and Meynet94 models. In all cases we see that the NOROT and Meynet94 models follow a very similar evolution. For the $U-B$ index the two are in good agreement; for the other indices the NOROT models are redder than the Meynet94 models from $\sim$4-6 Myr, then bluer from 6-7 Myr onward. The ROT models have a notably bluer $U-B$ index by $\sim$0.2-0.4 mag as compared to both non-rotating models, reflecting the brighter $U$ magnitude yielded by a rotating population; for the other colors, the ROT models are redder from 2-4 Myr, but bluer than both the NOROT and Meynet94 models at later ages. All of these variations are in agreement with the analyzes of V\'{a}zquez et al.\ (2007), who compared Starburst99 outputs adopting the Meynet94 models along with models by Meynet \& Maeder (2003) that include an early treatment of rotation with a consistent initial zero-age main sequence rotation velocity of $v_{ini} = 300$ km s$^{-1}$ across all masses. While the V\'{a}zquez et al.\ (2007) models extend to much later ages ($\sim$100 Myr), the behavior at these early ages is quite comparable even with the more detailed treatment of rotation effects in our ROT models.

\section{Discussion and Future Work}
We have used the Starburst99 evolutionary synthesis code to generate three stellar populations using different evolutionary tracks, including the recent tracks of Ekstr\"{o}m et al.\ (2012) that include a detailed treatment of stellar rotation effects. From comparisons of the resulting models, we have highlighted the extreme effects that rotation can have on massive star populations and their integrated properties. The treatment of rotation in the current models yields a substantial increase in the number of ionizing photons as well as the number of extremely luminous massive stars. Given the clear importance of rotation treatments in the physical properties predicted by evolutionary synthesis codes, we must now consider ways of refining stellar evolutionary models and their treatment of rotation if they are to be effectively utilized in future studies of extragalactic stellar populations.

Actual stellar populations are a mixture of stars with different initial velocities; however, the present models of Ekstr\"{o}m et al.\ (2012) use only one initial velocity, $v_{ini} = 0.4v_{crit}$, which provides a good fit to the observed mean velocities of B-type stars on the main sequence (Huang et al.\ 2010). There is, however, no guarantee that such a ratio can be similarly adopted for the whole mass range, particularly in the case of the most massive O-type stars. The Ekstr\"{o}m et al.\ (2012) tracks also model rotation in the theoretical frame described by Zahn (1992), which does not yet accommodate for effects such as the presence of surface or internal magnetic fields, both of which can also impact stellar rotation rates.Therefore, one possible improvement would be to compute population synthesis models that account for a realistic distribution of $v_{ini}$. Future work will also compare the Ekstr\"{o}m et al.\ (2012) tracks and our population synthesis models to the observed properties of local clusters. For example, Wolff et al.\ (2008) measured the massive star projected stellar rotation distribution in the LMC cluster R136. Comparing rotating evolutionary tracks and synthesis models at LMC metallicity to the emission line spectrum of this and other nearby clusters will allow us to test the models' predictions of individual stellar populations as well as the integrated cluster properties.

To investigate the potential effects of a population with a distribution of $v_{ini}$ values, we weighted a synthetic stellar population with 70\% rotating stars and 30\% non-rotating stars (see Figure 8). The resulting population shows a consistent decrease in the H I, He I, and He II ionizing continuums from $\sim$2 Myr onward while still producing more photons than either the NOROT or Meynet94 models. This suggests that such a mix of rotation rates may yield a less extreme increase in the hardness of the FUV spectrum. Our current models may overestimate the effects of a rotating stellar population due to neglecting these predicted effects of a velocity distribution. However, our present results do demonstrate the effect of a population of fast-rotating stars (or well-mixed stars following a nearly homogeneous evolution) on ionizing outputs. Depending on the frequency of such fast-rotating stars in the early Universe, the present results could reduce the number of stars required for reionization in the early universe.
 
In addition to different rotation rates, other elements of the evolutionary tracks such as mass loss and binarity can have profound effects on the resulting stellar populations. Georgy (2012) recently examined the effects of different mass loss treatments in the Geneva ROT models, finding that mass loss during the critical red supergiant phase can strongly impact the lifetime and terminal position of a star on the H-R diagram. Binary interactions can also profoundly affect many aspects of a star's evolution, including properties such as rotation-driven mixing and surface rotation velocities that have been shown here to be key in determining the physical properties and lifetimes of stars (de Mink et al.\ 2009, 2011). With a high predicted binary fraction for massive stars (e.g. Kobulnicky \& Fryer 2007), accommodating these effects could be key when modeling a massive star population. Finally, future Geneva evolutionary tracks at lower metallicities will also offer valuable information on how rotation impacts stars with lower natal abundances. Leitherer (2008) notes that the effects of rotation should become more pronounced as metallicity decreases.

Beyond adjustments to the stellar evolutionary models, future work will also consider other approaches to modeling and analyzing the integrated properties of rotating stellar populations. Possible updates to the temperature treatment and Wolf-Rayet definition currently adopted in Starburst99 will need to accommodate the new predictions of rotating evolutionary models, including the broader range of initial masses and ages for Wolf-Rayet stars and the effects of a rotating stellar interior on determinations of surface temperature. The evidence found here for more luminous main-sequence stars, corresponding to larger radii, could also impact the IMF assumptions made when modeling a rotating stellar population. Additionally, in this paper we have focused on examining the properties of a single coeval population of massive stars. Future models will examine how these rotation effects impact the properties of a simulated stellar population undergoing continuous star formation, widely considered to be a more effective means of modeling star formation in galaxies (e.g. Kewley et al.\ 2001, Fernandes et al.\ 2003, Izotov \& Thuan 2004, Levesque et al.\ 2010). Finally, by using the FUV spectra produced by Starburst99 in photoionization codes such as Mappings III (e.g. Sutherland \& Dopita 1993, Groves et al.\ 2004), we will be able to directly compare the resulting models to observed spectra of star-forming galaxies.

Previous stellar evolutionary models have included only cursory treatments of stellar rotation. However, from this work it is clear that rotation is a key component in modeling the physical properties of a stellar population. These in turn can have drastic effects on predictions of star-forming galaxies' spectral properties, and therefore on determinations of their stellar mass, star formation rate, and star formation history. With rotation treatments still being refined, and with predictions that these effects will be enhanced at lower metallicities, improvements in our modeling and understanding of stellar rotation effects are vital to the modeling and interpretation of star-forming galaxy samples out to high redshifts.

We thank the anonymous referee for valuable feedback on this manuscript. We also acknowledge useful correspondence with Charles Danforth, Cyril Georgy, and Mike Shull. EML is supported by NASA through Einstein Postdoctoral Fellowship grant number PF0-110075 awarded by the Chandra X-ray Center, which is operated by the Smithsonian Astrophysical Observatory for NASA under contract NAS8-03060. SE, GM, and DS are supported by the Swiss National Science Foundation.

\clearpage

\begin{figure}
\center
\includegraphics[width=9.5cm]{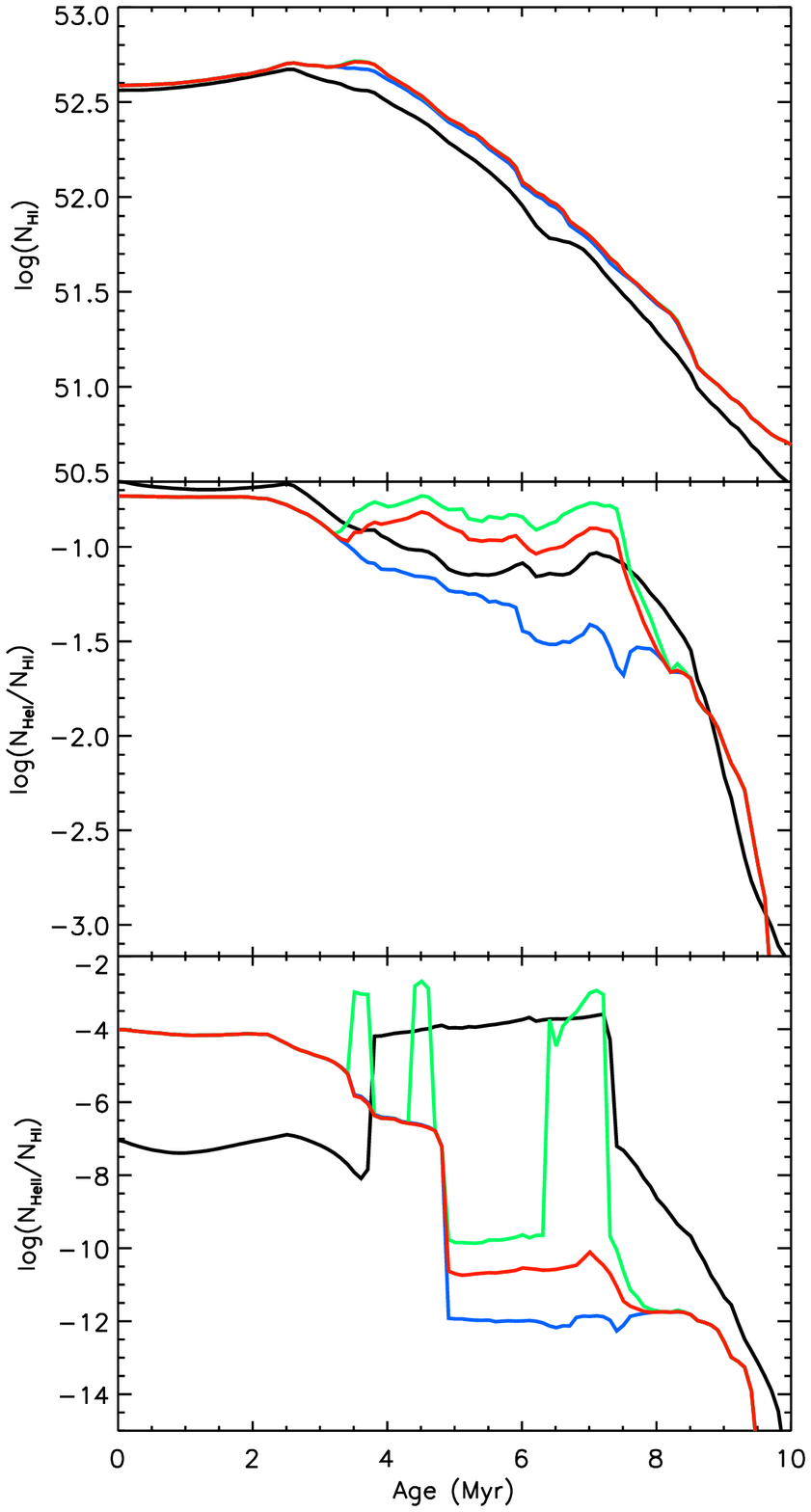}
\caption{Comparison of the H I (top), log(He I/H I) (center), and log(He II/H I) (bottom) photons produced by the Starburst99 code when adopting the ROT stellar evolutionary tracks and altering the atmosphere model treatments. These include the default treatment, adopting the extended Hillier \& Miller (1998) atmospheres with a weighting factor of 0.6 (red), as well as these same atmospheres with weighting factors of 1.0 (green) and 0.1 (blue) and the plane-parallel Lejeune et al.\ (1997) models with a default weighting factor of 1.0 (black). The weighting factor corresponds to the percent of stars in the synthetic population whose model atmospheres are computed using core temperatures (as opposed to surface temperatures) from the evolutionary tracks.}
\end{figure}

\begin{figure}
\center
\includegraphics[width=10cm]{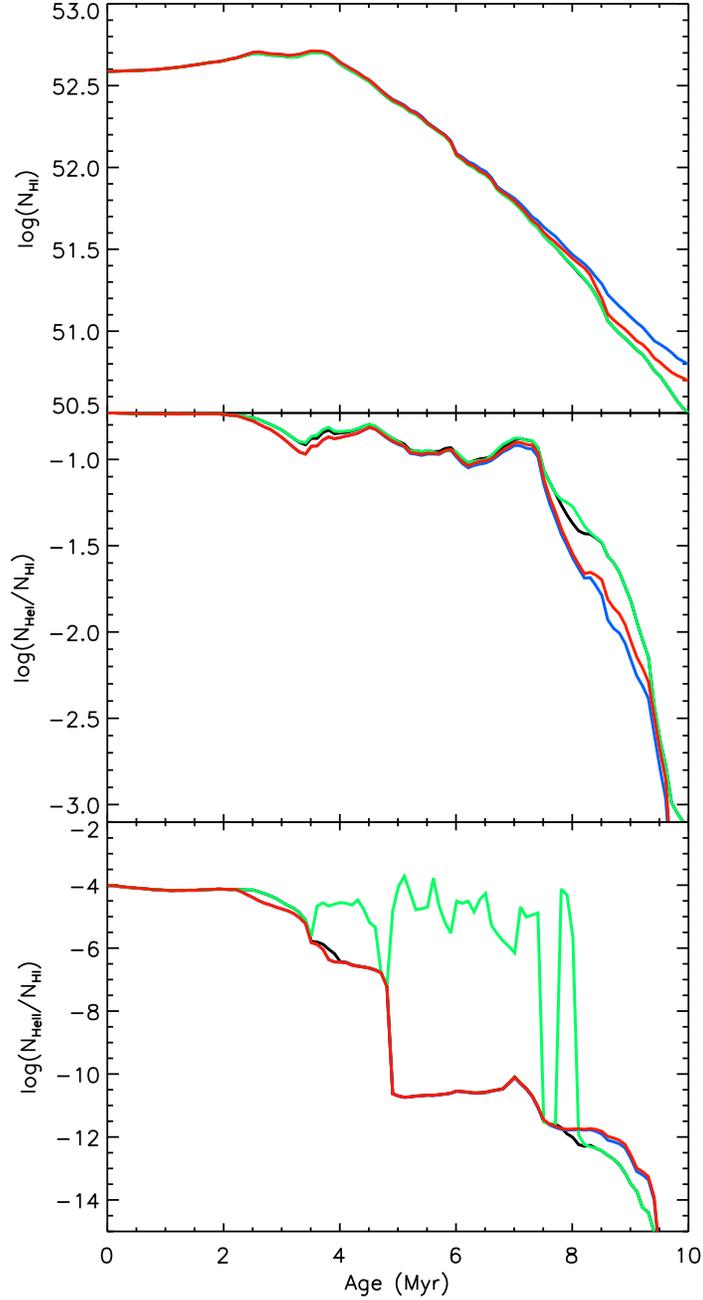}
\caption{Comparison of the H I (top), log(He I/H I) (center), and log(He II/H I) (bottom) photons produced by the Starburst99 code when adopting the ROT stellar evolutionary tracks and different criteria for the Wolf-Rayet star definition. These include the default criteria of log($T_{\rm eff}$) $> 4.4$ and a surface hydrogen fraction $X_H < 0.4$ (red), as well as log($T_{\rm eff}$) $> 4.8$ and $X_H < 0.4$ (green); log($T_{\rm eff}$) $> 4.0$ and $X_H < 0.4$ (blue); and log($T_{\rm eff}$) $> 4.4$ and $X_H < 0.1$ (black).}
\end{figure}

\begin{figure}
\center
\includegraphics[width=11cm]{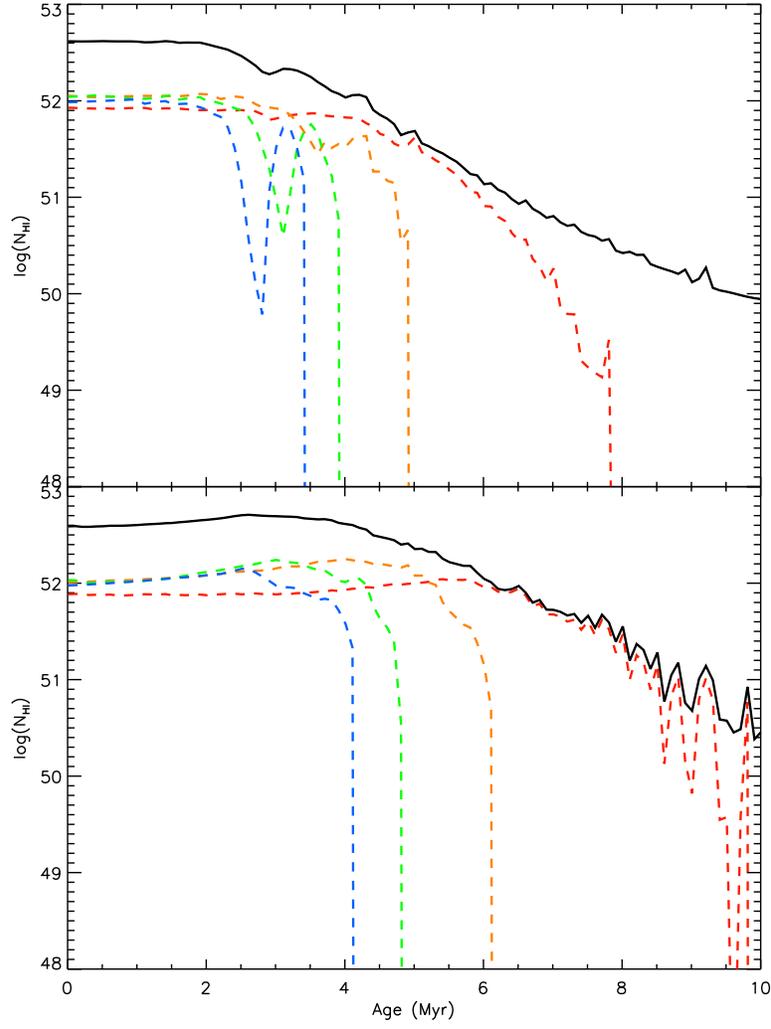}
\caption{Comparison of the contributed number of H I photons as a function of time for different mass intervals in the NOROT (top) and ROT (bottom) coeval stellar populations produced by the Starburst99 code. Intervals of 80$M_{\odot}$-100$M_{\odot}$ (dashed blue), 60$M_{\odot}$-80$M_{\odot}$ (dashed green), 40$M_{\odot}$-60$M_{\odot}$ (dashed orange), and 20$M_{\odot}$-40$M_{\odot}$ (dashed red) are illustrated as well as the total number of Lyman photons produced by the stellar population (solid black line).}
\end{figure}

\clearpage
\begin{figure}
\center
\includegraphics[width=11.6cm,angle=90]{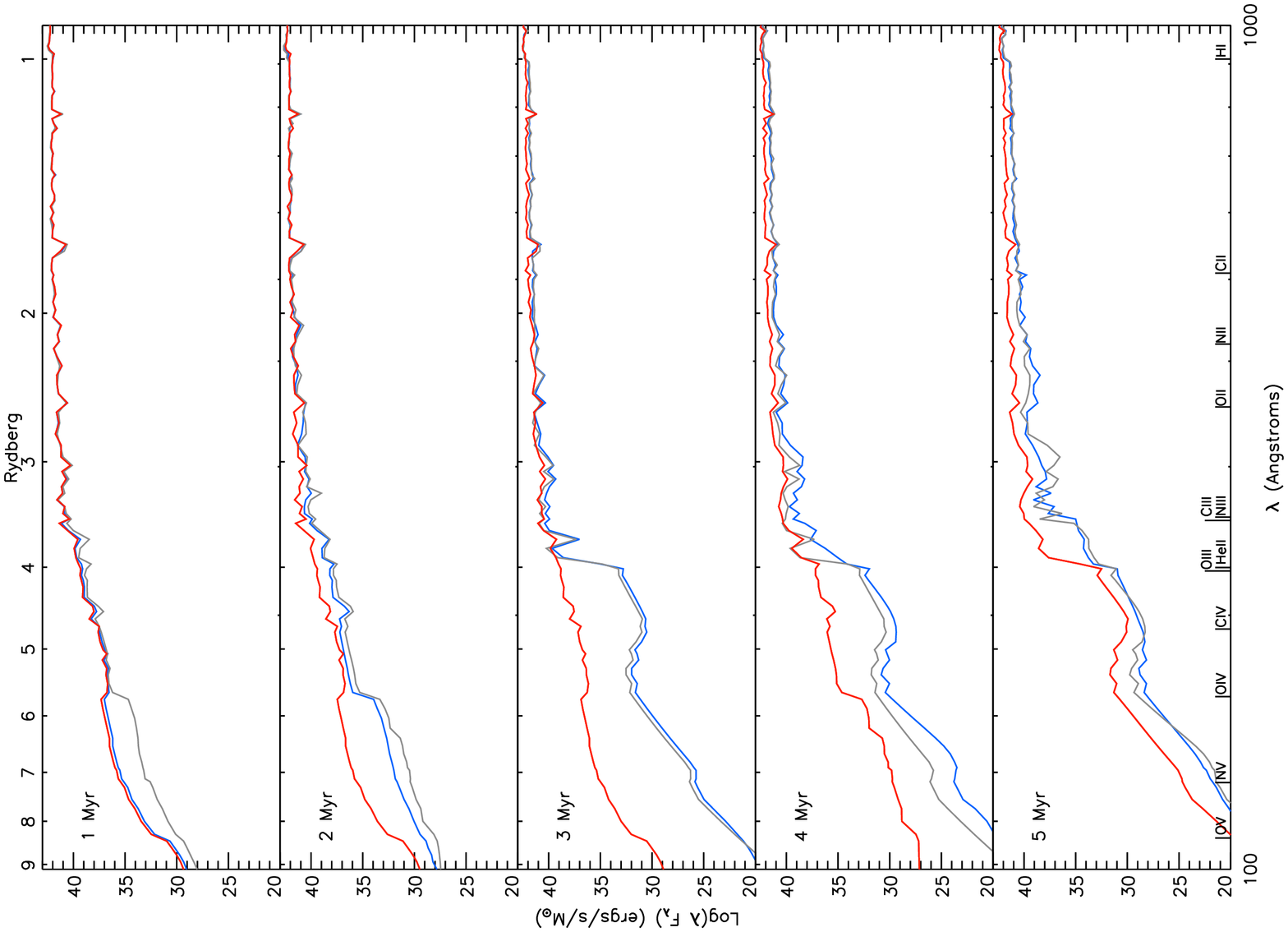} 
\includegraphics[width=11.6cm,angle=90]{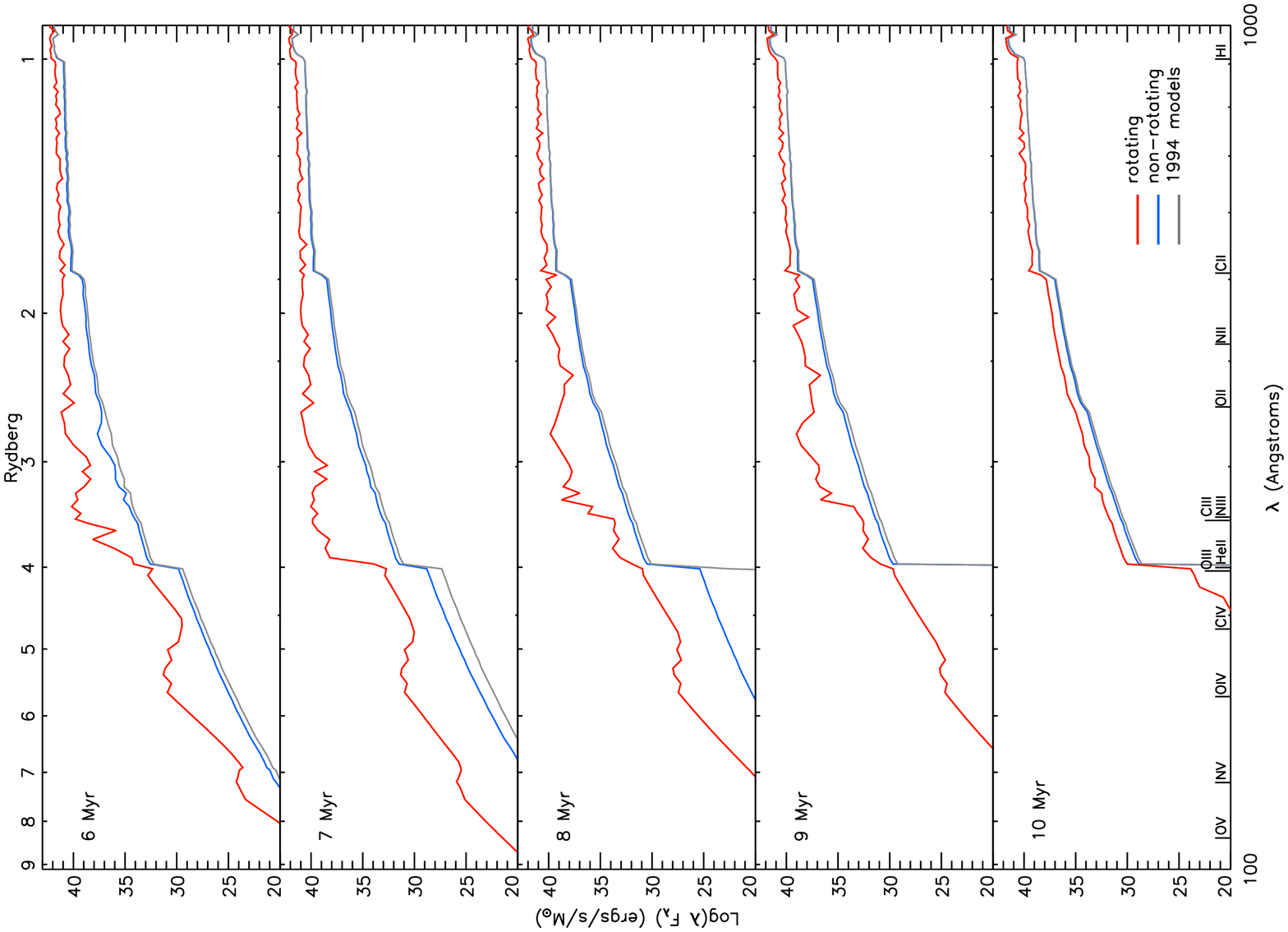} 
\caption{Synthetic FUV spectra generated using Starburst99 when adopting the ROT (red), NOROT (blue), and Meynet94 (gray) stellar evolutionary tracks, showing the evolution of the spectrum produced by a single coeval stellar population as it ages from 1 to 10 Myr. Key ionization potentials are marked on the x-axis.}
\end{figure}
\clearpage

\begin{figure}
\center
\includegraphics[width=11cm]{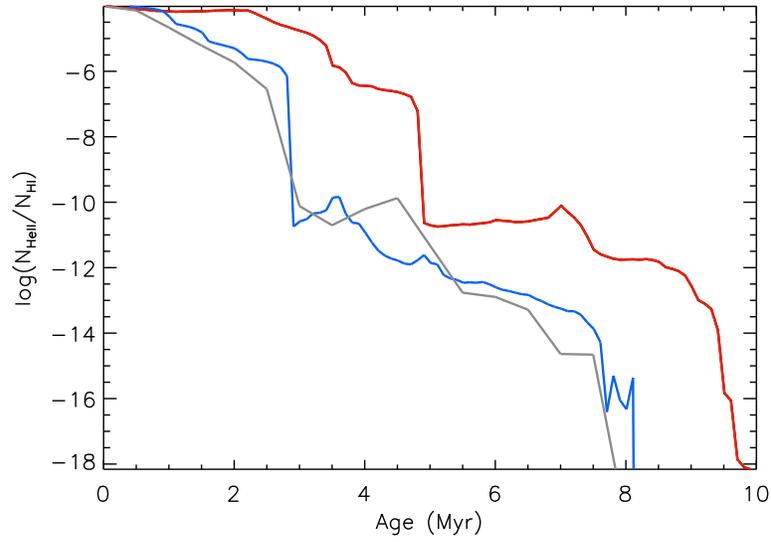}
\caption{As in Figure 4, but showing the evolution of the He II/H I continuum ratio with time.}
\end{figure}

\begin{figure}
\center
\includegraphics[width=11cm]{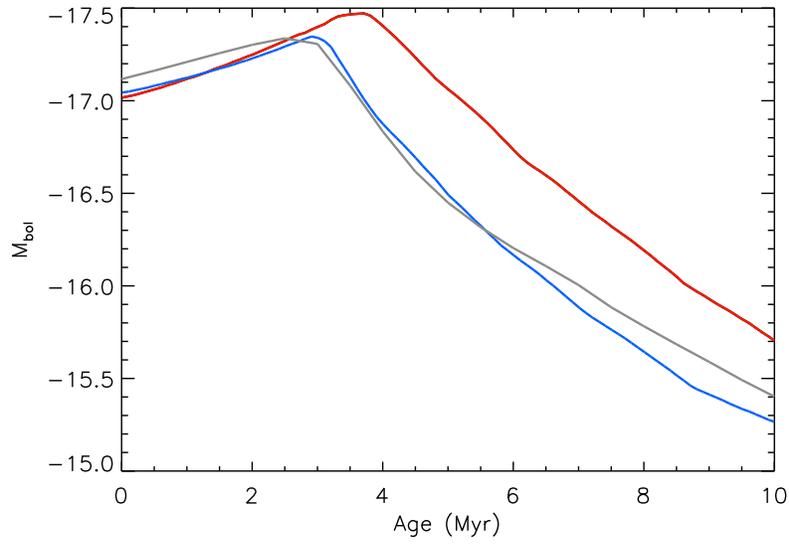}
\caption{As in Figure 4, but showing the evolution of the stellar population's $M_{\rm bol}$ with time.}
\end{figure}

\begin{figure}
\center
\includegraphics[width=8.0cm]{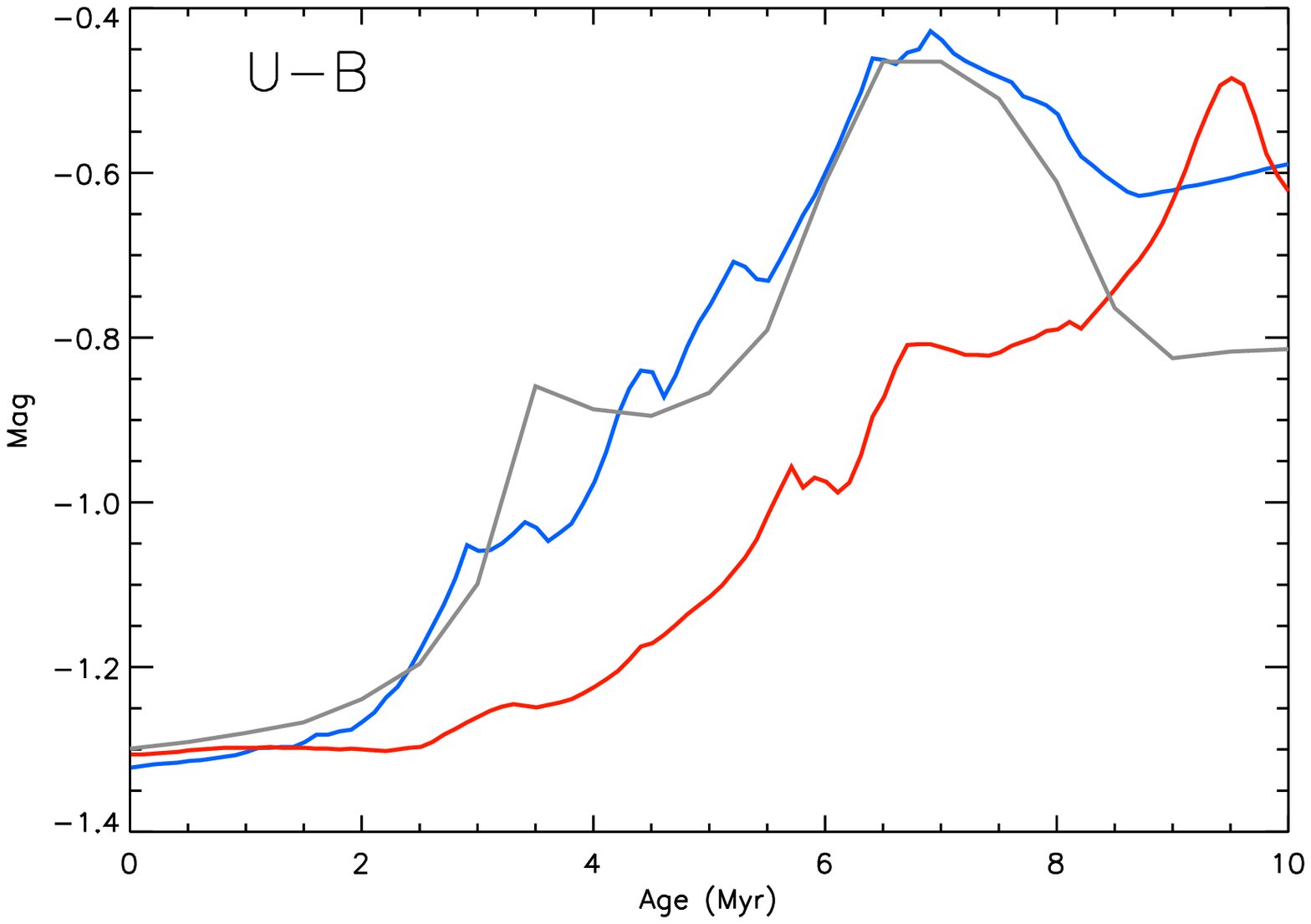}
\includegraphics[width=8.0cm]{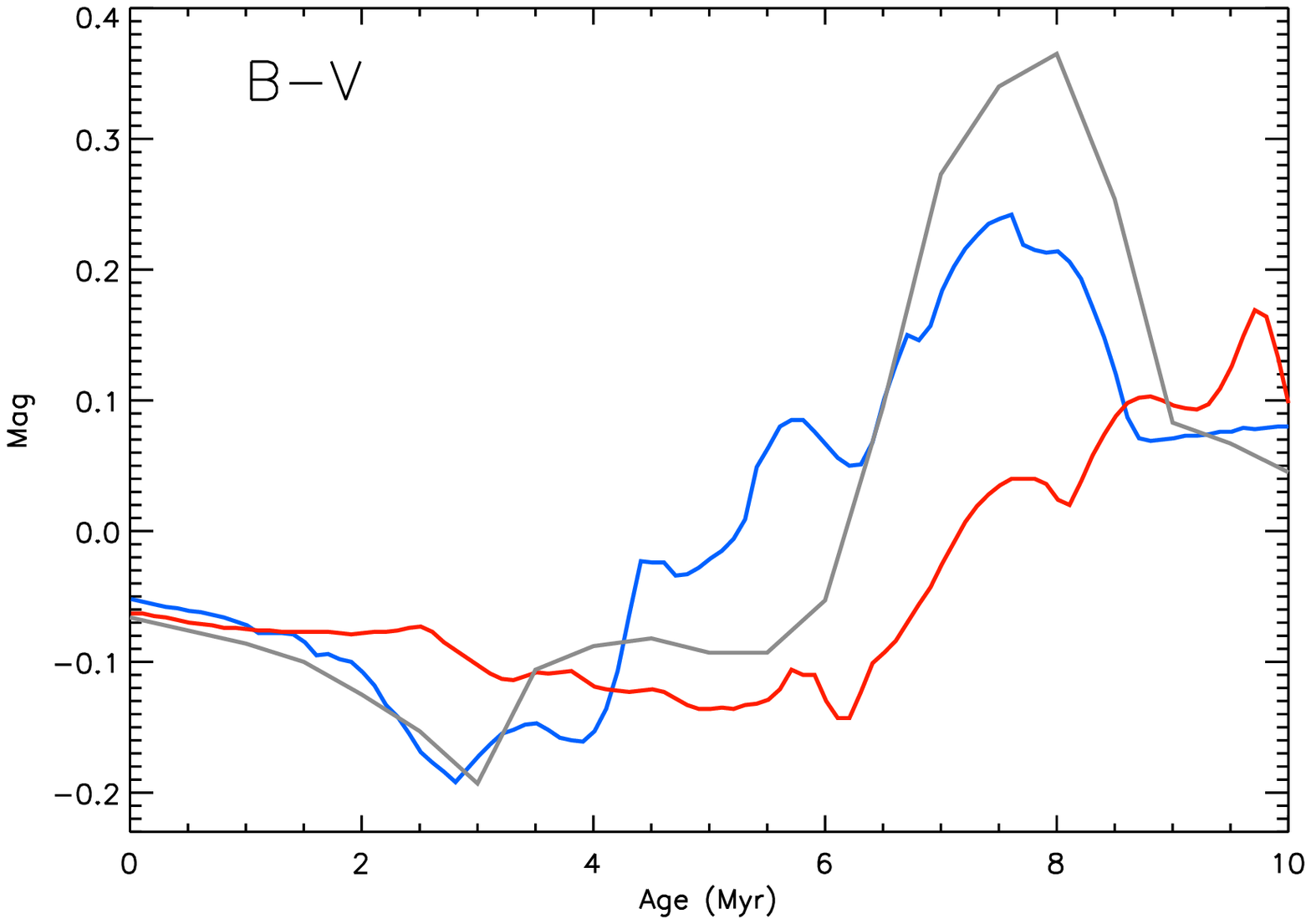}
\includegraphics[width=8.0cm]{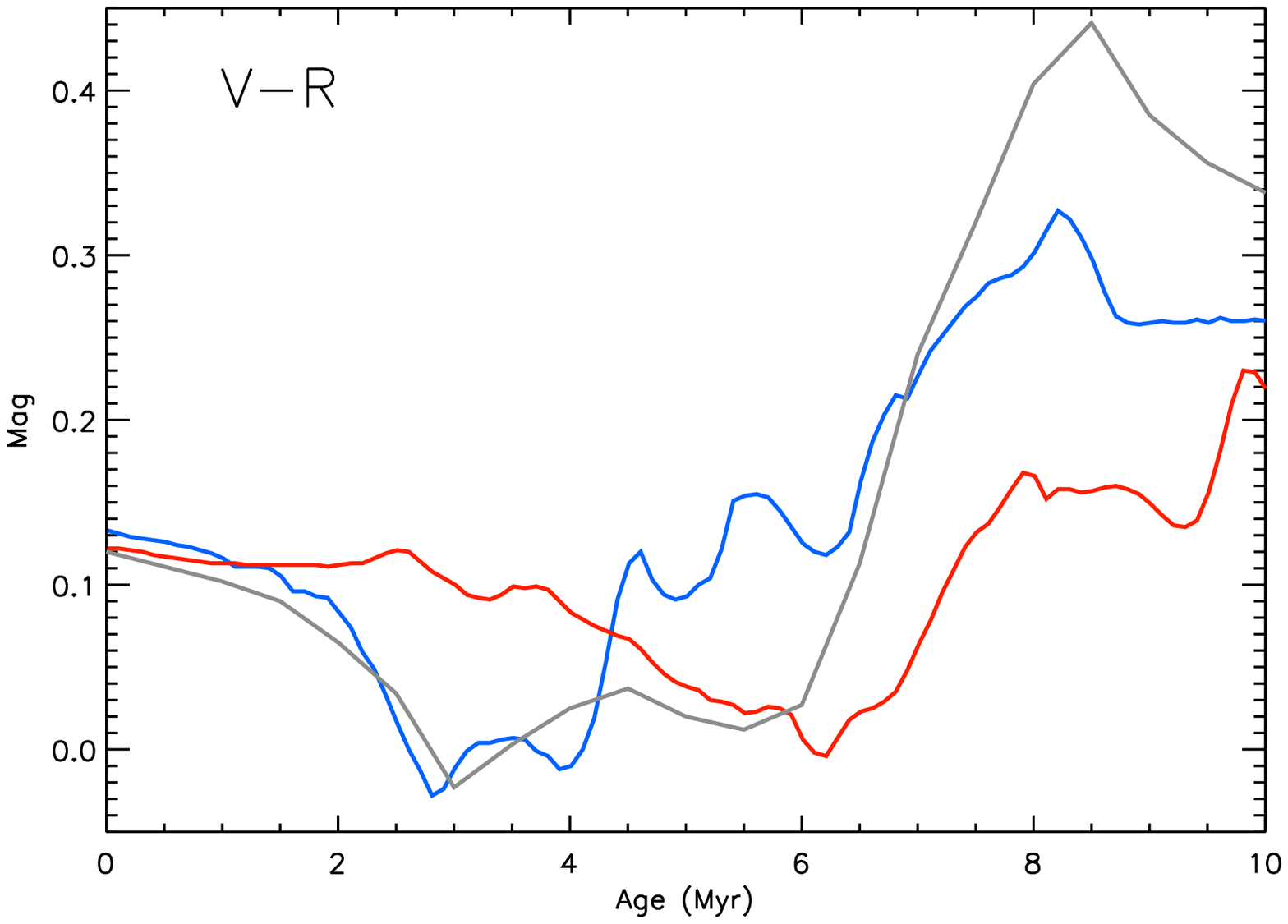}
\includegraphics[width=8.0cm]{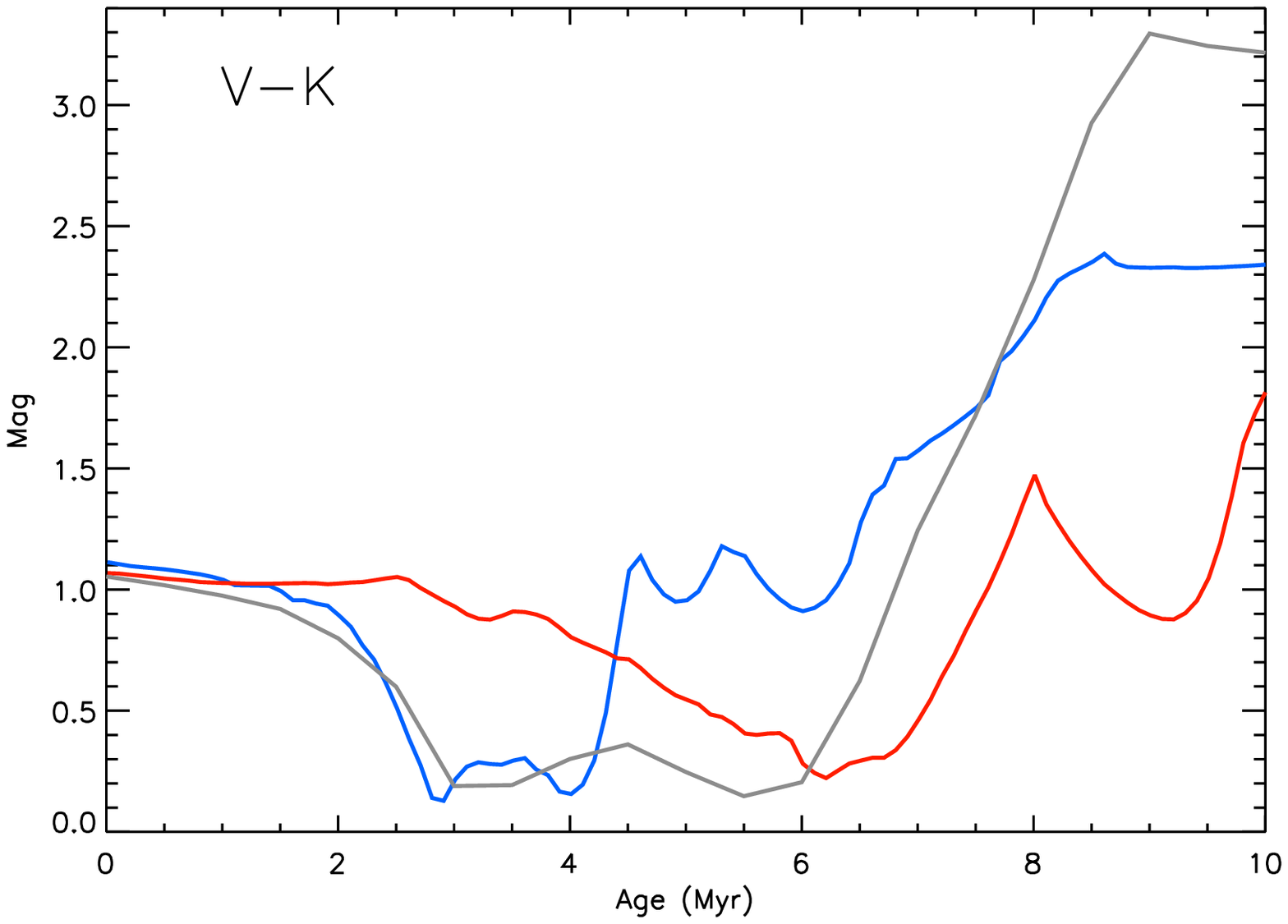}
\caption{As in Figure 4, but showing the evolution of the $U-B$ (top left), $B-V$ (top right), $V-R$ (bottom left), and $V-K$ (bottom right) colors with time.}
\end{figure}

\begin{figure}
\center
\includegraphics[width=10cm]{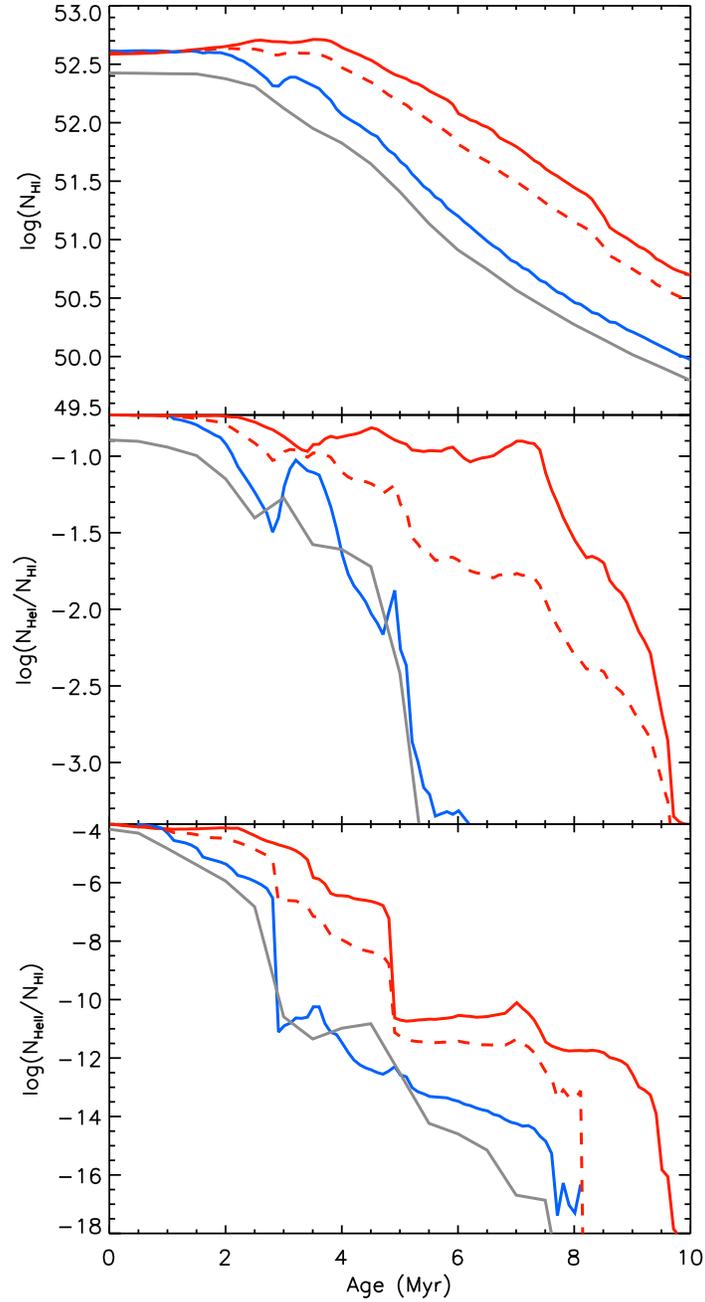}
\caption{Comparison of the H I (top), log(He I/H I) (center), and log(He II/H I) (bottom) photons produced by the Starburst99 code when adopting the NOROT (blue), Meynet94 (gray), and two treatments of the ROT (red) stellar evolutionary tracks. The solid red line adopts 100\% ROT evolutionary tracks, while the dashed line adopts a combination of 70\% ROT evolutionary tracks and 30\% NOROT evolutionary tracks.}
\end{figure}

\end{document}